\documentclass[runningheads]{svmult}

\usepackage{makeidx}   
\usepackage{graphicx}  
\usepackage{subeqnar}  
\usepackage{multicol}  
\usepackage{physprbb}  
\makeindex             

\begin{document}
\title*{ X-ray and radio bright type Ic SN 2002ap -- a hypernova without
an associated GRB} 
\toctitle{X-ray and radio bright SN 2002ap}
%
%
\titlerunning{X-ray and radio bright SN 2002ap}
%
\author{F. K. Sutaria\inst{1}
\and A. Ray\inst{2}
\and P. Chandra\inst{2,3}}
%
\authorrunning{F. K. Sutaria et al. }
%
%
\institute{The Open University, Milton Keynes, MK7 6AA, U. K.
\and Tata Institute of Fundamental Research, Bombay 400 005, India
\and Joint Astronomy Programme, Indian Institute of Science, Bangalore 560 012, India} 

\maketitle              

\begin{abstract}
Combined X-ray (0.3 -10 keV) and Radio (0.61 and 1.42 GHz)
observations of the type Ic 
SN 2002ap are used here, to determine the origins of the prompt
X-ray and Radio emission from this source. Our analysis of the XMM-Newton
observations suggests that the
prompt X-ray emission originates from
inverse Compton scattering of photospheric thermal emission
by energetic electrons. In addition, we use the reported
multifrequency VLA observations of this supernova.
We compare the early radiospheric
properties of SN 2002ap with those of SN 1998bw (type Ic) and SN 1993J
(type IIb), to contrast the prompt emission
from a GRB associated SN and other supernovae without such counterparts. 


\end{abstract}

\section{Introduction}
The nature and origins of X-ray and radio emission from supernovae
is a subject of much interest. This is mainly because early
(few days to years) emission in these bands occurs from a shocked,
dense CSM around the progenitor, and this can be used to infer the mass loss
rate and the nature of the progenitor. Further, early synchrotron radio
emission can also occur from relativistic electrons in  amplified magnetic
fields within the ejecta itself and its detection constrains theoretical
models on emission and reabsorption mechanisms in shocked matter.

SN 2002ap is an energetic ($E_{\rm explosion} \sim 4 - 10
\times 10^{51}$ ergs \cite{Maz02}), 
one of the closest (in M 74, distance $d= 7.3$ Mpc 
\cite{Sha96} ), type Ic SN
which exploded on Jan. $28 \pm 0.5 $, 2002 \cite{Maz02}. 
It was also one of the earliest detections in multiple bands, 
as in SN 1987A and SN 1993J.
Interest in highly energetic ("hypernova") type Ib/Ic SNe
has been intense after the temporal and positional near-coincidence 
of GRB 980425 with the type Ic SN 1998bw implied a 
possible GRB-SNe connection for the
long duration GRBs. 

We report our findings from the combined, near-simultaneous 
observations of SN 2002ap with XMM-Newton, GMRT \cite{Sutaria02} and 
VLA \cite{Ber02}. 
For a detailed analysis of the X-ray data and 
early 610 and 1420 MHz observations of SN 2002ap with GMRT (Giant 
Meterwave Radio Telescope), we refer the reader to \cite{Sutaria02}. 

\section{XMM-Newton and GMRT Observations and Analysis}

SN 2002ap was detected by XMM-Newton EPIC CCDs between 5.025
to 5.42 days after explosion \cite{Pas02} 
(See also \cite{Sor02}, \cite{Sutaria02}), with an effective exposure time
of $\sim 25.5$ ks. After subtracting the contribution 
from the nearby "contaminating"
X-ray source CXU J013623.4+154459 in the $50''$ spectral extraction region
on the EPIC-PN CCD, we estimate the 0.3-10 keV flux from SN 2002ap as 
$1.07^{+0.63}_{-0.31} \times 10^{-14}$ ergs cm$^{-2}$ s$^{-1}$.
Because the 
source is very faint, both thermal bremsstrahlung model 
($N_H= 0.49 \times 10^{22}$ cm$^{-2}$, $kT= 0.8$ eV) and
a simple powerlaw ($N_H=0.42 \times 10^{22}$ cm$^{-2}$, 
spectral index $\alpha =2.5$) fit the
sparse spectra well, with $\chi^2/d.o.f. = 1.2/20$. 
Simultaneous XMM-Newton Optical Monitor (OM) observations of M74 in the UVW1
band yield a flux of $ 7.667(\pm 0.002) \times 10^{-15}$ erg cm$^{-2}$ s$^{-1}$
$\AA^{-1}$.

%
%
%
%
%

SN 2002ap was observed with the Giant Meterwave Radio Telescope (GMRT)
at 0.61 GHz 8.56 days after explosion, and at 1.42 GHz 42 days after explosion.
The exposure time for each observation was 4 hours. We did not detect the SNe
in either observation and the upper limits on the flux have been tabulated in 
table \ref{tab: GMRT}.

\section{Results and Discussion}

The earliest radio detection of SN 2002ap was $\sim 4.5$ days after the 
explosion, in the 8.46 MHz band, and the frequency of the peak radio flux 
declined gradually from from 8.46 GHz to 1.43 GHz over a period of 10 days 
from the explosion epoch (VLA observation, \cite{Ber02}) -- a clear indication 
of the presence of relativistic e$^-$.
This wavelength dependence of the radio turn-on can be due to
either free-free absorption (FFA) in the thermally excited, 
homologously expanding matter overlying the 
interaction region, or synchrotron self-absorption
(SSA) by the relativistic $e^-$ responsible for radio emission, 
accelerated in regions 
close to the expanding interaction region \cite{Che01}.
 In the case of SN 2002ap, the combined GMRT upper limits \cite{Sutaria02} and
 VLA observations \cite{Ber02} are best fit by the SSA model 
(fig. \ref{fig:  GMRT+VLA})
in the optically thin, spherically symmetric limit, consistent with the early 
optical observations. The corresponding values of spectral 
index $\alpha$, magnetic 
field $B$, and the radius of the radiosphere $R_r$ are tabulated in Tab. 
\ref{tab: bestfit8.96}.   

\begin{table} 
\begin{minipage}[t]{60mm}
\caption{GMRT observations of SN 2002ap \label{tab: GMRT}}
\begin{tabular}{ccccc}
\\
\hline
Date     & $\nu$ & Resolution &  2$\sigma$ &  RMS \\
of       &       &            &  Flux      & mJy/     \\
Obs. & (MHz) & (arcsec)   &  mJy  &  beam \\
\hline
\\
5 Feb'02 & 610  & 9.5 x 6    & $<0.34$  & 0.17 \\
8 Apr'02 & 1420 & 8 x 3     & $ <0.18$ & 0.09 \\
\\
\hline
\end{tabular}
\end{minipage} 
\ \
\begin{minipage}[t]{58mm}
\begin{center}
\caption{ Best Fit parameters of the SSA Model
for SN 2002ap on day 8.96. $F_{p}$ is the peak flux
emitted at frequency $\nu_{p}$, $B$ is the ambient 
magnetic field in the radiosphere
at radius $R_r$.  \label{tab: bestfit8.96}}
\begin{tabular}{ccccc}
\\
\hline
$\alpha$ & $\nu_{p}$ & $ F_{p}$ & $R_{r}$ & B  \\
         & GHz       & $\mu$ Jy   & cm. & G \\

\hline
\\
0.8  & 2.45  & 397  &  $3.5\times 10^{15}$ & 0.29 \\
\hline
\end{tabular}
\end{center}
\end{minipage}
\end{table}


The early Radio and X-ray turn-ons imply that the mass-loss rate
of the progenitor was relatively low, at $\dot M \le 6 \times 10^{-5}$
$M_{\odot}$ yr$^{-1}$ (\cite{Sutaria02}, using VLA detection by \cite{Ber02},
and the SSA models for X-ray absorption by \cite{Che94} ).

\subsection{Origins of X-ray emission}

Using the optical observations of SN 2002ap \cite{Mei02} on the same
epoch as the XMM-Newton observation, we find that the radius of the optical
photosphere $R_{opt.} = 3.4 \times 10^{14}$ cm. 
We note that the radiosphere in the SN (table 2) was well outside this optical 
photosphere at a similar epoch. However, Compton optical depth considerations 
(\cite{Sutaria02}, \cite{Fra82}) suggest that X-ray production region is closer 
to the optical photosphere.

 Free-free emission mechanisms are likely to be dominated by Compton
cooling. The high ejecta velocity ($ v \ge 20,500$ km s$^{-1}$ on day 3.5),
coupled with high implied 
temperature of the shocked circumstellar medium ($ T \sim
10^9$ K) and a limited cool absorbing shell suggests that the high energy
photons would have a flat tail, upto $\sim 100$ keV. By contrast, 
the X-ray spectra
is quite soft, with thermal bremsstrahlung $T_{B} = 0.8$ keV. Hence, free-free
absorption is unlikely to explain the observed X-ray spectra. However,
Compton cooling is the dominant radiative mechanism when the optical
photospheric temperature is $T_{eff} \ge 10^4  K$. Using the Monte Carlo
simulations of Compton scattered spectra for the shocked CSM \cite{Poz77},
we find that the observed flux by XMM can be accounted for both in terms
of energetics and spectrum if the electron plasma has $T_e$ $\sim$ few 
times of $10^9$ K (See \cite{Sutaria02}).
Table \ref{tab: progenitor} gives  the properties of Comptonised plasma for
two likely progenitor systems (\cite{Lan99}, \cite{Hab02}) for this event.

\begin{table}
\begin{minipage}[t]{55mm}
\caption{ Comptonising Plasma Properties at t = 5d
with x-ray energy index $\gamma=3 $ and $R_{opt} = 3.4 \times 10^{14}$ cm
\label{tab: progenitor}}
\begin{tabular}{ccccc}
\\
\hline
\bf Scenario & $\dot{M}_{-5}$  & $u_{w1}$  & $\tau_{e}$ &  $T_{e}$  \\
             & $\times 10^{-5}$ & 10       &            &           \\
        & $M_{\odot}$ yr$^{-1}$  & km/s & $\times 10^{-4}$ & $10^{9} K$ \\
\hline
Wolf-  & 1.5  & 58   & $ 4 $ & $2$ \\
Rayet  &      &      &       &     \\
\hline

Case-BB    & 10 & 58 & $25$ & $1.5$ \\
Binary  &    &    &       &		\\ 

\hline
\end{tabular}
\end{minipage} \ \
\begin{minipage}[t]{60mm}
\caption{ Least-squares fitted SSA equipartition parameters for 
SN 1993J, SN 1998bw and 
SN 2002ap $\sim 11$ days after explosion. 
\label{tab: 3SNe}}
\small
\begin{tabular}{ccccccc}
\hline
SNe    &$\nu_{p}$ & $F_{p}$ & $\theta_{eq}$ & $U_{eq}$  & $B_{0}$ & $R_{0}$ \\
       &          &         &          & $\times 10^{44}$ &      & $\times 10^{15}$ \\
       &  GHz     & mJy     & $\mu$ as      & ergs      & G       & cm      \\
\hline
2002ap & 2.45  & 0.48  & 39.0  & 6.9 & 0.47    & 4.80 \\
1998bw & 5.5   & 50.4  & 112.4 & 35000 & 0.23    & 68.4 \\
1993J  & 30.5  & 22.3  & 17.6  & 5.0 & 3.54    & 1.08 \\
\hline
\end{tabular}
\end{minipage}
\end{table}

\begin{figure}[h]
\begin{minipage}[t]{6.5cm}
\caption{Best fit synchrotron self-absorption spectrum (solid line)
for SN 2002ap radio emission on day 8.96 with energy spectral index = -0.8.
Also shown is the free-free absorption model (dashed line) with
parameters used taken from \cite{Ber02}. The VLA data is marked with
a ``$\times$", while the GMRT upper limit at 610 MHz is marked by a ``$\bigcirc$
".}
\label{fig:  GMRT+VLA}
\resizebox{\hsize}{!}{\includegraphics*[1.9cm,5.7cm][18.9cm,23.4cm]{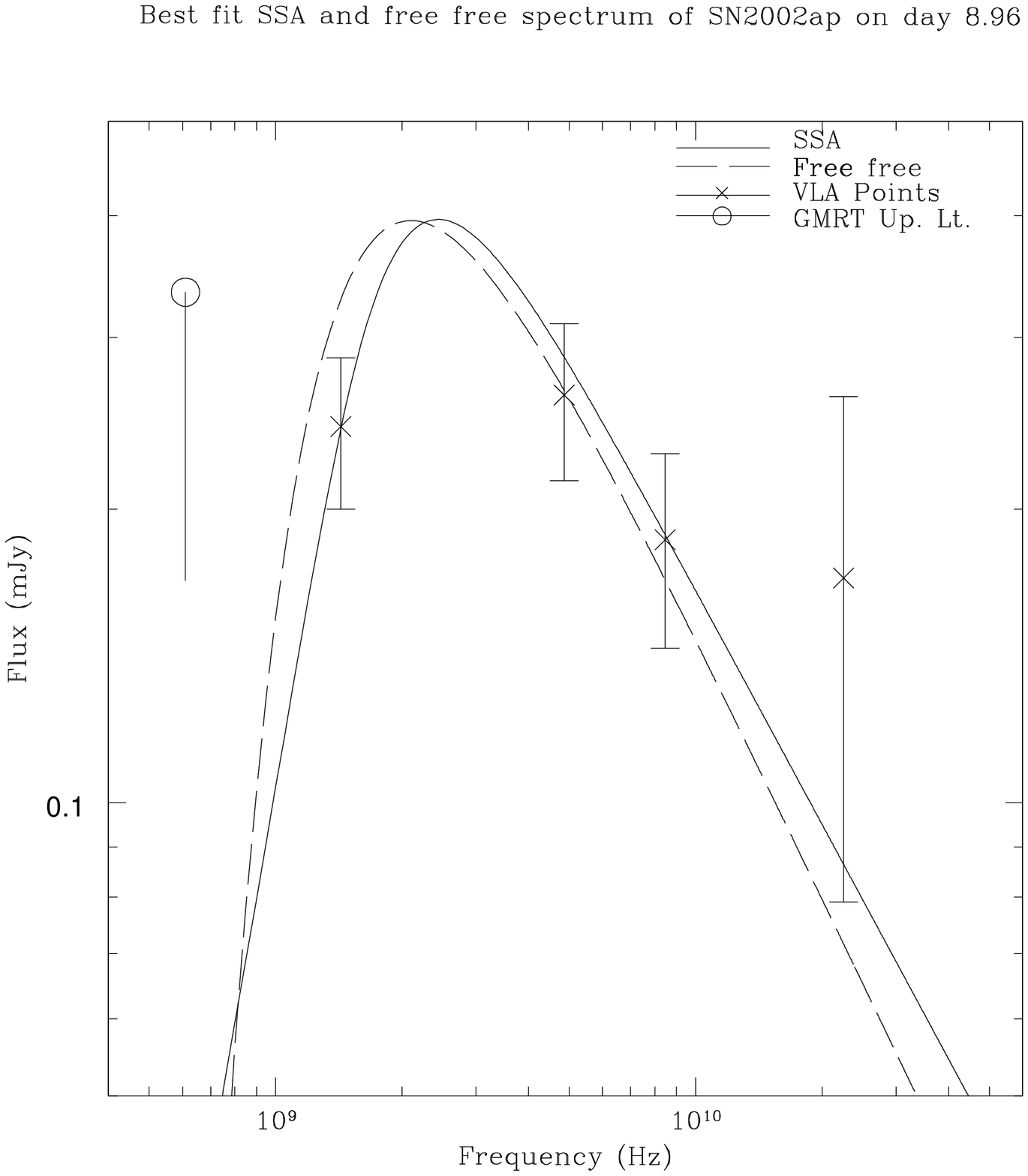}}
\end{minipage} \ \
\begin{minipage}[t]{6.5cm}
\caption{Comparison of spectra of three SNe (SN 1993J \cite{vanDyk94},
SN 1998bw \cite{Kul98}  and SN 2002ap \cite{Ber02}
near day 11 after explosion.
Solid lines show the best fit synchrotron self absorption
model and dashed lines show the corresponding free free absorption model.}
\label{fig: 3SNe}
\vskip 0.8 true cm
\resizebox{\hsize}{!}{\includegraphics*[1.9cm,5.7cm][19.0cm,23.3cm]{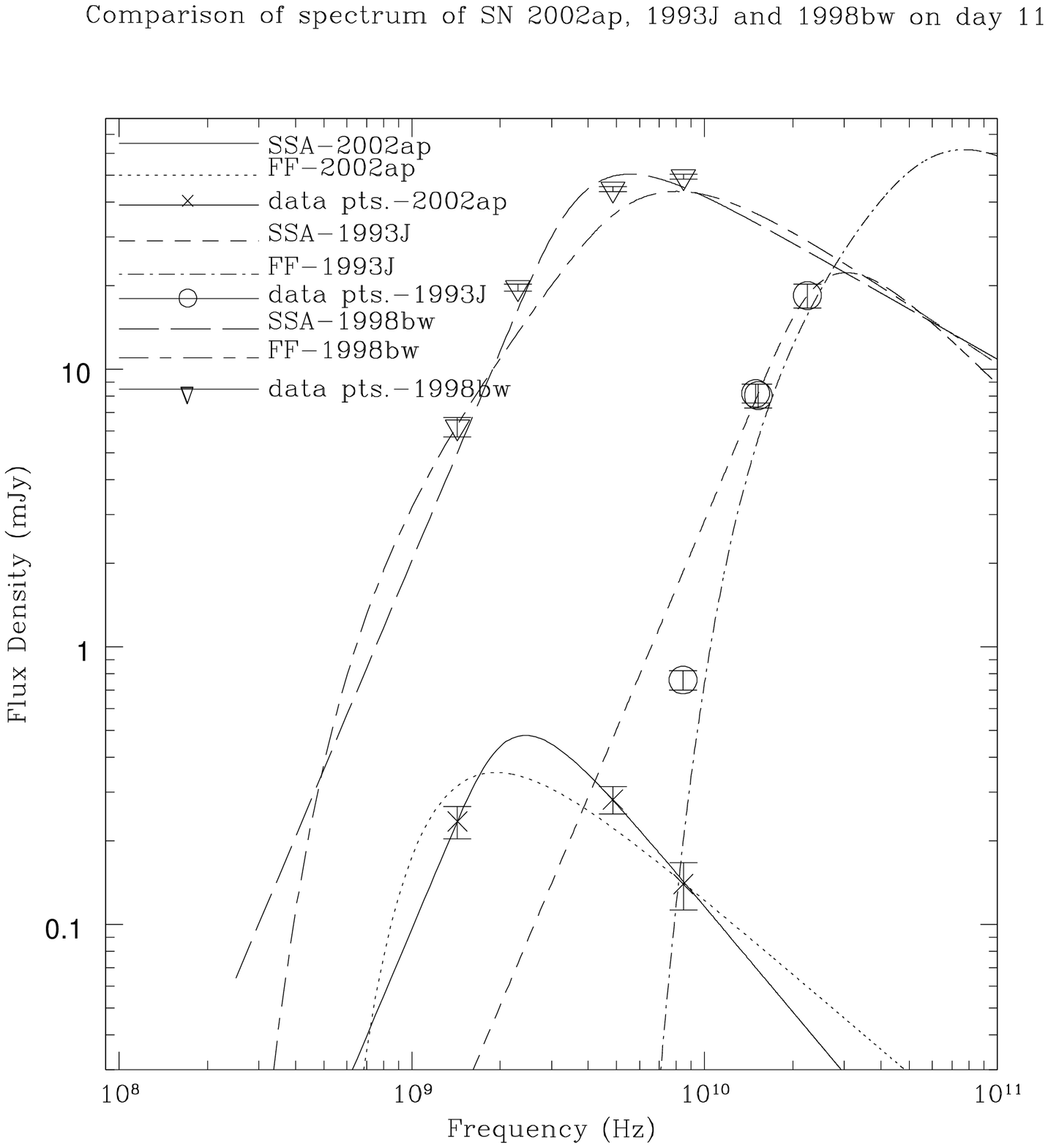}}
\end{minipage}
\end{figure}

Finally, we compare the rarely observed early 
radio spectra of SN 2002ap with that of SN 1993J (type IIb) and SN 1998bw (type Ic,
GRB-association), 11 days after explosion. 
In Fig. \ref{fig: 3SNe}, we fitted the SSA model to the data, and in 
table \ref{tab: 3SNe}, we estimate the energy $U_{eq}$ in the radiating plasma
and the magnetic field $B_{0}$, from equipartition arguments, and determine 
corresponding the angular radius $\theta_{eq}$  of the
radio-sphere. Comparing $\theta_{eq}$ for the 3 SNe, the hydrodynamic 
shock in SN 1998bw, causes more rapid expansion leading to larger
radio-sphere, while SN 2002ap is similar to SN 1993J across type 
classification, irrespective of presence or absence of H- and 
He-envelope in the progenitor. 

We thank the XMM-Newton helpdesk and the staff of GMRT,
especially S. Bhatnagar, for many useful communications during
data processing.

%

\end{document}